# How to induce multiple delays in coupled chaotic oscillators?

Sourav K. Bhowmick[1], Dibakar Ghosh[2], Prodyot K.Roy[3], Jürgen Kurths[4,5], Syamal K. Dana[1]

[1] *CSIR-Indian Institute of Chemical Biology, Jadavpur, Kolkata* 700032, *India.*
[2] *Physics and Applied Mathematics Unit, Indian Statistical Institute, Kolkata-700108, India.*
[3] *Department of Physics, Presidency University, Kolkata* 700073, *India.*
[4] *Potsdam Institute for Climate Impact Research, 14473 Potsdam, Germany*
[5] *Institute for Physics, Humboldt University, 12489 Berlin, Germany*

Lag synchronization is a basic phenomenon in mismatched coupled systems, delay coupled systems and time-delayed systems. It is characterized by a lag configuration that establishes a unique time shift between all the state variables of the coupled systems. In this report, an attempt is made how to induce multiple lag configurations in coupled systems when different pairs of state variables attain different time shift. A design of coupling is presented to realize this multiple lag synchronization. Numerical illustration is given using examples of the Rössler system and the slow-fast Hindmarsh-Rose neuron model. The multiple lag scenario is physically realized in an electronic circuit of two Sprott systems.

**Lead Paragraph**

**Lag synchronization (LS) is usually studied in instantaneously coupled mismatched oscillators, in the presence of coupling delay in identical systems and time-delayed systems. LS is characterized by one unique lag time or time shift that is established between all the pairs of state variables of the coupled oscillator while the amplitudes remain strongly correlated. The amount of lag time, of course, can be varied by tuning the parameter mismatch or the delay in the coupling function as the case may be. The characteristic lag time or delay between the coupled oscillators may be used as a form of transmitting information. A question is raised if it is possible to induce multiple lags in coupled systems, chaotic or periodic. Transmitting different delays through different pairs of state variables of two unidirectionally coupled oscillators (periodic or chaotic) was not reported so far, to our best knowledge. If implemented, a simultaneous transmission of multiple delays as information bits can be of advantage for communication systems. To achieve this goal, a design of coupling approach is presented here to explain how to induce multiple lag or delay in two unidirectionally coupled chaotic oscillators with numerical examples and electronic circuit experiment.**

## I. Introduction

Lag synchronization (LS)[1] as observed in instantaneously mismatched chaotic oscillators, delay coupled oscillators or delay coupled time-delayed systems has one and unique lag configuration. The amplitudes of all the pairs of state variables of the coupled oscillators remain strongly correlated but shifted by a common time or delay. LS is, particularly, considered as an important scenario in coupled mismatched oscillators since the complete synchronization (CS)[2] is an ideal case and not usually observable in practical systems. CS defines a state of exact correlation in both the amplitude and the phase which is only possible if two oscillators are identical. But never two systems can be exactly identical in nature or engineering. In practical systems, an almost CS is thus seen for large or strong coupling in closely identical systems. Interestingly, although the parameter mismatch plays destructive effect on the CS, the amplitudes show to remain strongly correlated for a weaker coupling while the state variables of the coupled systems are shifted by a constant time. The amount of time shift or lag time is determined by the amount of mismatch and the coupling strength. Such a LS scenario is seen above a critical coupling[3] which is for sure smaller than the critical coupling for CS in identical oscillators. For a given parameter mismatch, the time shift is unique for all the pairs of state variables of the coupled systems and decreases with coupling strength above the critical value and becomes almost zero for large coupling when the coupled systems emerge into an almost CS state. This lag configuration has potential applications in transmitting information[4] in unidirectionally coupled oscillators. Alternatively, a time delay is used[5] in the coupling function of two chaotic oscillators to realize a LS scenario. As a result, a delay or lag time can be considered as information transmitted from a driver to a response oscillator.  An almost identical but a delayed version of the driver signal is retrieved at the response system. The retrieved time shift or delay may be considered as the information bit.

In this paper, we treat the problem whether it is possible to transmit different delays through different pairs of state variables of two coupled oscillators. In other words, if different delays are used in a vector type coupling in different state variables, is it possible to establish stable LS with different lag configurations or time shift in different pairs of state variables? It may help transmit separate information simultaneously using different driver variables. In the past, a kind of multiple lag configuration is reported in the context of intermittent lag synchronization[6] when two coupled systems switches between different lag configurations, however, different pairs of state variables still follow one unique lag configuration at any duration of time. To our best knowledge, no attempt is so far made to realize multiple lag configurations or multiple LS (MLS) either in periodic or chaotic system.

In real complex networks such as brain[7], stock-market[8], signals travel simultaneously between individual nodes through multiple paths with different lag or delays. On the other hand, multiplexing[9] is a usual form of transmitting digital signals by time sharing of a common communication channel. It will be an added practical advantage for communication application if it is possible to transmit several information signals in terms of multiple delays simultaneously via different state variables of two oscillators. In the perspective of neuronal networks too, it is well known[10] that one neuron receives information from other neurons at different time instants since they travel different path length to arrive at the destination neuron. In this context, we make an attempt to engineer a coupling scheme that can induce separate delays in different state variables of two drive-response type coupled systems. This is manifested as the response variables being shifted with different time lag from the corresponding driver variables. To our best search of literature, a co-existence of dual-lag is found[11] that considers two semiconductor lasers which are coupled optically via two paths of different lengths. However, no general coupling strategy is proposed that can be implemented to dynamical system, in general. We address this issue of MLS in chaotic oscillators using a design of a coupling scheme based on the Hurwitz matrix stability[12]. The important feature of the method is that the coupling is assumed to be unknown *a priori*. Given the model of a dynamical system, a desired state of synchronization is first targeted and then the coupling function is derived to realize the targeted stable state using a general measure of stability. We illustrate the theory with numerical examples of the Hindmarsh-Rose model[13] and the Rössler system [14]. Furthermore, we implement the MLS in an electronic experiment using a coupled Sprott system[15].

The rest of the paper is organized as follows: the theory of the MLS in chaotic systems is discussed in section II. In section III, numerical examples of MLS are presented using mismatched Rössler system and identical Hindmarsh-Ross model. Experimental observation of MLS is described using a Sprott circuit in section IV. Our results are summarized in section V.

## II. Design of coupling for MLS

Consider a chaotic system as driver with parameter mismatch,

$$\dot{y} = f(y,\eta) + \Delta(y,\eta), \qquad y \in R^n \qquad (1)$$

where $\Delta f(y,\eta) = f(y, \eta + \Delta \eta) - f(y,\eta)$ contains the mismatched terms, in general, where $\eta$ is a vector of system parameters. Otherwise if all the parameters appear in the linear term of $f(.)$, the mismatch term is more simplified, $\Delta f(y,\eta) = f(y, \Delta \eta)$.

Next consider another system

$$\dot{x} = f(x, \eta), \qquad x \in R^n \qquad (2)$$

and we target a goal dynamics $x(t) = g_\tau = y_\tau$ as a desired response, when $y_\tau = [y_1(t-\tau_1), y_2(t-\tau_2), y_3(t-\tau_3), ......., y_n(t-\tau_n)]^T$ and $\tau_i \geq 0 \, (i=1,2,...,n)$ are coupling delays and $T$ denotes transpose of a matrix. The response system after coupling is

$$\dot{x} = f(x, \eta) + D(x, g_\tau) \qquad (3)$$

where the delay coupling term $D(x, g_\tau)$ is defined by

$$D(x, g_\tau) = \dot{g}_\tau - f(g_\tau, \eta) + \left( H - \frac{\partial f(g_\tau)}{\partial g_\tau} \right)(x - g_\tau) \qquad (4)$$

$\frac{\partial f}{\partial g_\tau}$ is the *Jacobian* and $H$ is an arbitrary constant $n \times n$ matrix. The error signal of the coupled system is defined by $e = x - g_\tau$. Using Taylor series expansion, $f(x)$ can be written as

$$f(x) = f(g_\tau + e) + \frac{\partial f(g_\tau)}{\partial g_\tau} e + ..... \qquad (5)$$

Restricting to the first order term, the error dynamics can be easily obtained[12] as $\dot{e} = He$ using (3)-(5). This ensures that $e \rightarrow 0$ for $t \rightarrow \infty$ if $H$ is a Hurwitz matrix whose eigenvalues all have negative real parts and when asymptotically stable LS is obtained. If $\tau_1 = \tau_2 = ... = \tau_n$, the conventional LS scenario is seen[1] but alternatively, MLS is observed for $\tau_1 \neq \tau_2 \neq .... \neq \tau_n$. This is possible since the stability condition does not involve the delay time as elaborated later. The $H$- matrix is constructed from the *Jacobian* $\partial f(g_\tau)/\partial g_\tau$. If a system is known, the *Jacobian* is derived by a linearization of the system. H-matrix is then constructed using a set of rules: the elements of a *Jacobian* matrix that contain state variables are replaced by a set of constant $p_i$ keeping other elements (zero or constant) unchanged. Once the H-matrix is formed, its characteristic equation is derived and the Routh-Hurwitz (RH) criterion[16] is applied to obtain the condition for which all eigenvalues have negative real parts. As an example of a 3D system, the characteristic equation of the H-matrix is given by $\lambda^3 + a_1 \lambda^2 + a_2 \lambda + a_3 = 0$ where the coefficients $a_1$, $a_2$, $a_3$ are defined by the elements of the matrix, i.e. the system parameters and $p_i$. Next, apply the Routh-Hurwitz (RH) criterion as given by $a_1 > 0, a_3 > 0, a_1 a_2 > a_3$ that confirms all eigenvalues of $H$ have negative real parts and $H$ becomes Hurwitz and it ensures stability of the error dynamics ($e$) at zero. The stability of a desired synchronized state is thereby established. From these conditions, for a given set parameter values, the range of $p_i$ values is determined for which the RH criterion is valid. The system parameters decide the dynamics (periodic or chaotic) of the coupled system which remains undisturbed by the choice of the $p_i$ values. It is not difficult now to implement the method, in numerical simulations, once the Hurwitz matrix is designed by the appropriate choice of its $p_i$ parameters from the given range of values. To complete the design of the coupling term $D(x, g_\tau)$, the coupling delay $\tau$ is to be decided next which can be arbitrarily chosen as desired without any loss of stability of MLS. The stability condition of MLS depends on $p_i$ only for a set of system parameters and is independent of the coupling delay $\tau$. This is a great advantage in obtaining the stable MLS state. The coupling design is illustrated in the next section with numerical examples.

## III. Numerical examples

Next, we illustrate the design of coupling based on Hurwitz matrix stability for MLS in identical as well as mismatch systems. We show that the stability of MLS is maintained even in the presence of a parameter mismatch. We take paradigmatic examples, the Rössler system and the Hindmarsh-Rose neuron model.

### A. Mismatched oscillator: Rössler system

We start with two mismatched Rössler oscillators and show that mismatch does not affect the stability of MLS. We consider the unidirectional delay coupling when the driver is,

$$\dot{y} = f(y) + \Delta f(y) \quad \text{where } y = (y_1, y_2, y_3) \in R^3, \tag{6}$$

$$f(y) = \begin{pmatrix} -y_2 - y_3 \\ y_1 + by_2 \\ c + y_3(y_1 - d) \end{pmatrix} \quad \text{and} \quad \Delta f(y) = \begin{pmatrix} 0 \\ \Delta b y_2 \\ 0 \end{pmatrix}$$

and $\Delta b$ is a parameter mismatch. The *Jacobian* of the system and the H-matrix are,

$$J = \frac{\partial f}{\partial y} = \begin{bmatrix} 0 & -1 & -1 \\ 1 & b & 0 \\ y_3 & 0 & y_1 - d \end{bmatrix} \quad ; \quad H = \begin{bmatrix} 0 & -1 & -1 \\ 1 & b & 0 \\ p_1 & 0 & p_2 - d \end{bmatrix}$$

We construct the *H* from the above *Jacobian* matrix by following the rules as stated above. In our simulation we choose $p_1=5$, $p_2=-5$ when the eigenvalues of the H-matrix are -14.6605, -0.0947±0.9813i. As mentioned above, this choice is not a unique one, in fact, a wider choice[11] is available and the coupled system remains in the chaotic dynamics. It is, particularly, to mention that the choice ensures the chaotic regimes. After adding the coupling (2), the response system becomes,

$$\dot{x} = f(x) + D(x, y_\tau) \tag{7}$$

$$\text{where} \quad D = \begin{pmatrix} -y_2(t-\tau_1) - y_3(t-\tau_2) + y_2(t-\tau_2) + y_3(t-\tau_3) \\ -y_1(t-\tau_1) + y_1(t-\tau_2) + \Delta b y_2(t-\tau_2) \\ (p_1 - y_3(t-\tau_1))(x_1 - y_1(t-\tau_1)) + (p_2 - y_1(t-\tau_1))(x_3 - y_3(t-\tau_1)) + \\ y_3(t-\tau_3)(y_1(t-\tau_3) - d) - y_3(t-\tau_3)(y_1(t-\tau_1) - d) \end{pmatrix}$$

The targeted MLS manifold is,

$$x_1(t) = y_1(t-\tau_1), \quad x_2(t) = y_2(t-\tau_2), \quad x_3(t) = y_3(t-\tau_3).$$

In our simulations, we consider the coupling delays $\tau_1=1.0$, $\tau_2=2.0$, $\tau_3=3.0$, however, we can make any arbitrary choice (besides integer values) of the delays without disturbing the stability of the MLS. Figure 1(a) shows the time series of the driving signal $y_1(t)$ (blue line) and the response signal $x_1(t)$ (red line) for a lag time τ=1.0. To confirm the lag configuration, we estimate a similarity measure $S_{y_1x_1}$ between the variables ($y_1$, $x_1$) as shown in Fig. 1(b). The similarity measure[1] between two state variables $x_1(t)$ and $y_1(t)$ is defined as,

$$S_{y_1x_1} = \frac{\langle [x_1(t) - y_1(t-\tau_s)]^2 \rangle}{\sqrt{\langle x_1^2(t) \rangle \langle y_1^2(t) \rangle}} \tag{8}$$

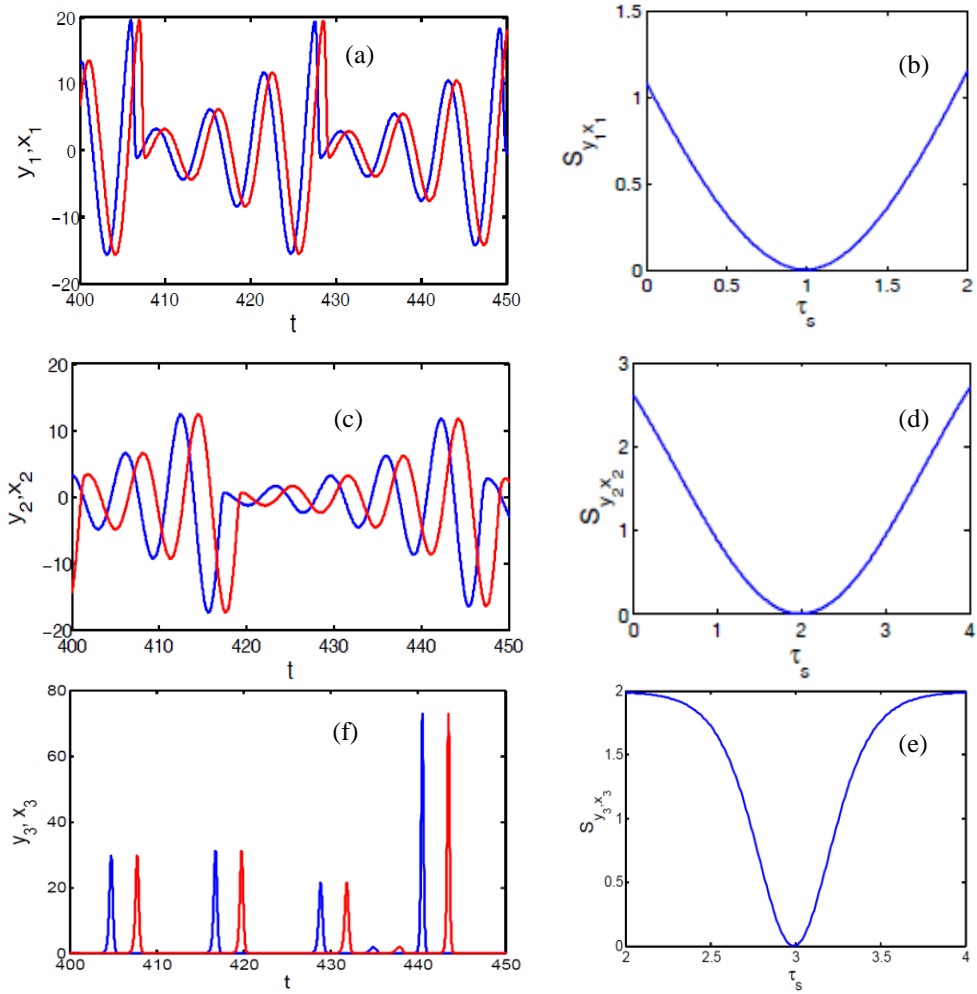

FIG. 1: (Color online) Multiple lag synchronization in coupled Rössler system in Eq.(6)-(7): (a) time series of $x_1(t)$(red line) and $y_1(t)$(blue line) shows lag synchronization, (b) using the similarity measure $S_{yx}$ in eq.(8) between ($x_1$, $y_1$), $\tau_s$=1.0, (c) time series of $x_2(t)$ and $y_2(t)$, (d) similarity measure between ($x_2$, $y_2$), $\tau_s$=2.0, (e) time series of $x_3(t)$ and $y_3(t)$, and (f) similarity measure between ($x_3$, $y_3$), $\tau_s$=1.0. The system parameters are $p_1$=5, $p_2$=-5, $b$=0.15, $c$=0.2, $d$=10 and $\Delta b$=0.05.

The $S_{y_1x_1}$ plot with an arbitrary $\tau_s$ shows a global minimum at zero for $\tau_s = \tau_{s0}$, which is the principal lag characteristic or configuration of the pair of time series. $S_{y_1x_1}$ has a global minimum closely at zero at $\tau_1 = \tau_s = 1.0$ (dimensionless) which estimates the lag time between the time series $y_1$ and $x_1$ in Fig. 1(a). Similarly Figs. 1(c) and 1(e) show the time series ($y_2$, $x_2$) and ($y_3$, $x_3$) with time lag 2.0 and 3.0 respectively. Figures 1(d) and 1(f) confirm the corresponding similarity measures of the pairs of time series ($y_2$, $x_2$) and ($y_3$,

$x_3$) where $S_{y_2 x_2}$ and $S_{y_3 x_3}$ also have global minima at zero at $\tau_s=2.0$ and $\tau_s=3.0$ respectively. Our numerical results confirm the MLS scenario where three different delays are introduced separately in the coupling function and each of them are exactly retrieved from the response signals. It is important to emphasize that the presence of a parameter mismatch in the coupled systems does not destabilize the MLS.

### B. Identical oscillator: Hindmarsh-Rose model

Next, we check the coupling scheme for MLS in a slow-fast system. For this, we choose the spiking-bursting Hindmarsh-Rose neuron model,

$$\dot{y} = f(y) \tag{9}$$

where $y = (y_1, y_2, y_3) \in R^3$ and $f(y) = \begin{pmatrix} y_2 - ay_1^3 + by_1^2 - y_3 + I \\ c - dy_1^2 - y_2 \\ r\{s(y_1 + 1.6) - y_3\} \end{pmatrix}$

where $y_1$ is the fast membrane voltage and, $y_2$ and $y_3$ are associated with fast and slow membrane currents, and $I$ is the bias current. For the numerical simulations, we take the system parameters as $a=1.0$, $b=3.0$, $c=1.0$, $d=5.0$, $s=5.0$, $r=0.003$, $I=4.1$. Now we assume to have an identical response system. As usual, the first step in the process of designing the coupling is to derive the *Jacobian* of the given model system and then to construct the H-matrix and to convert it into a Hurwitz by appropriate choice of its $p_i$ parameters as discussed above and a wider choice of it is available for this system too. Finally, we use eq. (4) to define the coupling for one given model system with a choice of system parameters of a desired dynamics (chaotic or periodic). For the Hindmarsh-Rose system, the *Jacobian* ($J$) and the corresponding H-matrix are given by,

$$J = \frac{\partial f}{\partial y} = \begin{bmatrix} -3ay_1^2 + 2by_1 & 1 & -1 \\ -2dy_1 & -1 & 0 \\ rs & 0 & -r \end{bmatrix} \text{ and } H = \begin{bmatrix} p_1 & 1 & -1 \\ p_2 & -1 & 0 \\ rs & 0 & -r \end{bmatrix}$$

The condition for $H$ to be a Hurwitz is derived as $p_1 < r+1$ for $p_2=0$ using the RH criterion[16] which obviously provides a wider choice of $p_i$ values. We decide our target to realize MLS in the response Hindmarsh-Rose system,

$$\dot{x} = f(x) + D(x, g_\tau), \quad x = (x_1, x_2, x_3) \in R^3 \tag{8}$$

where the goal dynamics is set as,

$$g_\tau = \begin{bmatrix} y_1(t - \tau_1) \\ y_2(t - \tau_2) \\ y_3(t - \tau_3) \end{bmatrix} = \begin{bmatrix} x_1(t) \\ x_2(t) \\ x_3(t) \end{bmatrix}$$

We make an additional approximation, $\tau_1 = \tau_3$, for simplification of the coupling when the slow response variable ($x_3$) will attain a lag identical to one of the fast response variables ($x_1$). We must mention that it does not affect the MLS scenario if $\tau_1 \neq \tau_3$. However, we make the approximation for numerical ease. Accordingly, the response system after coupling is derived using eq. (2) as given by,

$$\text{where } D = \begin{pmatrix} y_2(t-\tau_1) - y_2(t-\tau_2) + [p_1 + 3ay_1^2(t-\tau_1) - 2by_1(t-\tau_1)](x_1 - y_1(t-\tau_1)) \\ dy_1^2(t-\tau_1) - dy_1^2(t-\tau_2) + [p_2 + 2dy_1(t-\tau_1)](x_1 - y_1(t-\tau_1)) \\ 0 \end{pmatrix}$$

(9)

The condition for stability of MLS now depends only on the external parameter $p_1$ ($p_2$=0). MLS emerges when the coupling delays are targeted as non-identical. In the simulations, we consider $p_1$= -15, $p_2$=0 and coupling delays $\tau_1 = \tau_3 = 10$, $\tau_2 = 5$. This makes two different delays one each for the fast variables. Figure 2(a) shows the time series of the driving signal $y_1(t)$ (blue color) and the response signal $x_1(t)$ (red color) with a lag time $\tau_1 = 10$. Figure 2(b) shows the time series of ($y_2$, $x_2$) with a time lag $\tau_2 = 5$. To confirm LS, we calculate the similarity measure between pairs of variables ($y_1$, $x_1$), ($y_2$, $x_2$) as shown in Figs. 2(c) and 2(d). Figure 2(c) plots $S_{y_1 x_1}$ from the pair of time series ($y_1$, $x_1$) in Fig. 2(a), which has a global minimum near zero at $\tau_1$=$\tau_{s0}$=10.0 and confirm LS of $\tau_1$=10.0. Similarly, Fig.2(d) shows the similarity measure $S_{y_2 x_2}$ between the pair of state variable ($x_2$, $y_2$) in Fig. 2(b) and shows a global minimum at zero for $\tau_2$=$\tau_{s0}$=5.0.

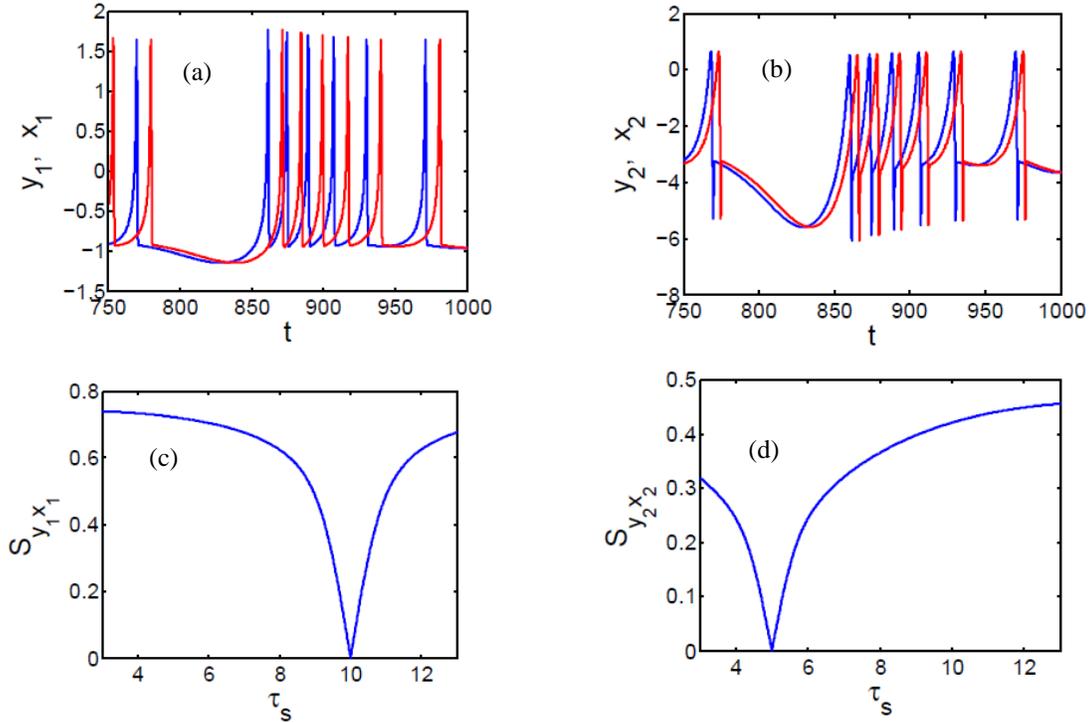

FIG. 2: (Color online) Multi-delay synchronization in Hindmarsh-Rose neural model: Time series of (a) ($y_1$, $x_1$) for $\tau_1$=10 (b) ($y_2$, $x_2$) for $\tau_1$=5. Similarity measure between ($y_1$, $x_1$) to confirm LS in (c), $\tau_s$ =10.0 and between ($y_2$, $x_2$) in (d) to confirm LS, $\tau_s$ =5.0.

The time series of the pair of slow variables are not shown here, however, they emerge into the same LS scenario as of the first pair of variables. The coupling design is thus able to induce two different LS configurations as targeted in this slow-fast system. Numerically, of course, we can also target a set of three delays for the two systems which is redundant here. We are interested to show the applicability of the coupling design for more than one lag configuration in the coupled system.

## V. Experimental observation

Finally, we physically implement the MLS scheme in an electronic circuit using a Sprott system,

$$\dot{y} = f(y) \qquad (10)$$

$$\text{where } y = (y_1, y_2, y_3) \in R^3, \quad f(y) = \begin{pmatrix} -ay_2 \\ y_1 + y_3 \\ y_1 + y_2^2 - y_3 \end{pmatrix}$$

For reducing complexity in the circuit implementation, we again consider two separate time delays in the coupling instead of three by setting $\tau_2 = \tau_3$. Our target is show that multiple delays can really be induced in two coupled systems. After coupling the response system becomes,

$$\dot{x} = f(y) + D \qquad (11)$$

$$\text{where } D = \begin{pmatrix} -ay_2(t-\tau_1) + ay_2(t-\tau_2) \\ y_1(t-\tau_2) - y_1(t-\tau_1) \\ y_1(t-\tau_2) - y_1(t-\tau_1) + [p_1 - 2y_2(t-\tau_2)](x_2 - y_2(t-\tau_2)) \end{pmatrix}$$

and $H = [0\ -a\ 0; 1\ 0\ 1; 1\ p_1\ -1]^T$.

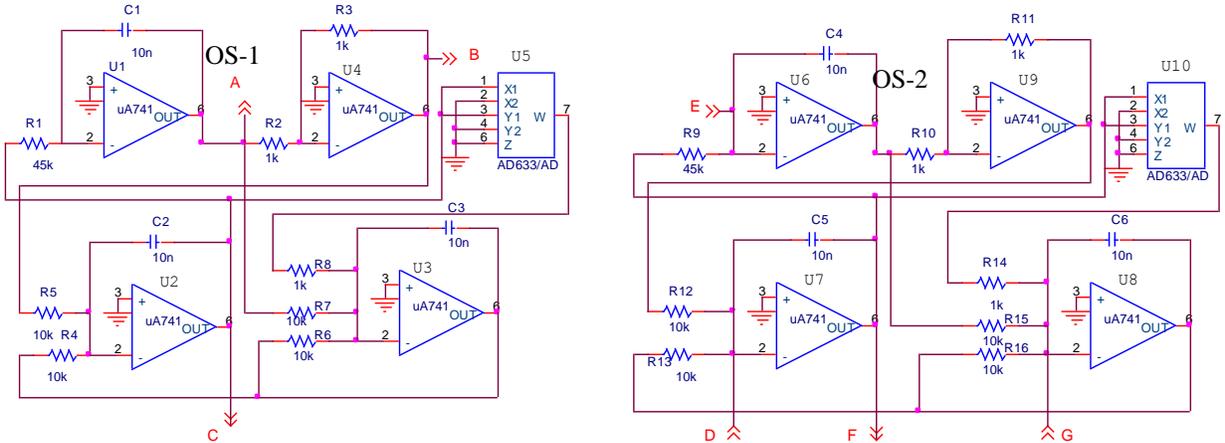

FIG. 3: Circuit of two Sprott systems: driver (left) and response (right) systems. Incoming and out-going connecting nodes (A-G) appropriately connect the delay couplers.

A physical realization of the uncoupled Sprott systems (10) and (11) is shown in Fig. 3 with circuit diagrams. The driver Sprott circuit (OS-1) is designed using three Op-amp (U1-U3) as integrators with associated resistances, capacitors and an inverting amplifier (U4); the multiplier U5 simulates the quadratic nonlinearity in the driver. Similarly the response circuit (OS-2) is designed using three integrators (U6-U8), one inverting amplifier (U9) and a multiplier U10. The delay coupler in Fig. 4 is designed using three op-amp (U11-U13), two multiplier (U14-U15) and associated resistance shown. The delay line in the coupler is designed[16] using a ladder LC network preceded by an isolating amplifier and followed by a non-inverting amplifier. This amplifier is used to compensate the attenuation in the signal due to leakage resistance in the inductors of the LC arrays. The lag time is now increased algebraically by simply adding one after another LC circuit in series as desired. The power supply for all active devices is ±9 Volt. The variables $y_1(t)$ and $x_1(t)$ of

(10) and (11) are recorded as output voltage of U1 and U6 respectively using a 2-channel digital oscilloscope (Tektronix TDS 2012B, 100MHz, 1GS/s) as shown in the lower row of Fig. 5. We find that the oscilloscope pictures of the driver and the response variables are in MLS for two arbitrary time lags or delays ($\tau_1$=150μs, $\tau_2=\tau_3$= 400μs) as designed by adding one after another LC circuit in the coupler. MLS between (10) and (11) is also investigated using numerical simulation. In simulation, we have considered $p_1$= -1, $\tau_1$=1 and $\tau_2$=3. Note that $p_1<1$ satisfies the RH condition. The numerical time series of ($y_1$, $x_1$), ($y_2$, $x_2$) and ($y_3$, $x_3$) are shown in the upper row of Fig. 5. Clearly two different time delays appear in three pairs of time series of the driver and the response system since we induce only two delays in the coupling function. A similar result is seen in the experimental results of three pairs of time series shown again in the lower panels (oscilloscope pictures). The pairs of time series ($y_2$, $x_2$) and ($y_3$, $x_3$) are shown in the middle and right panel in the lower row respectively as measured output of the (U2, U7) and (U3-U8) respectively.

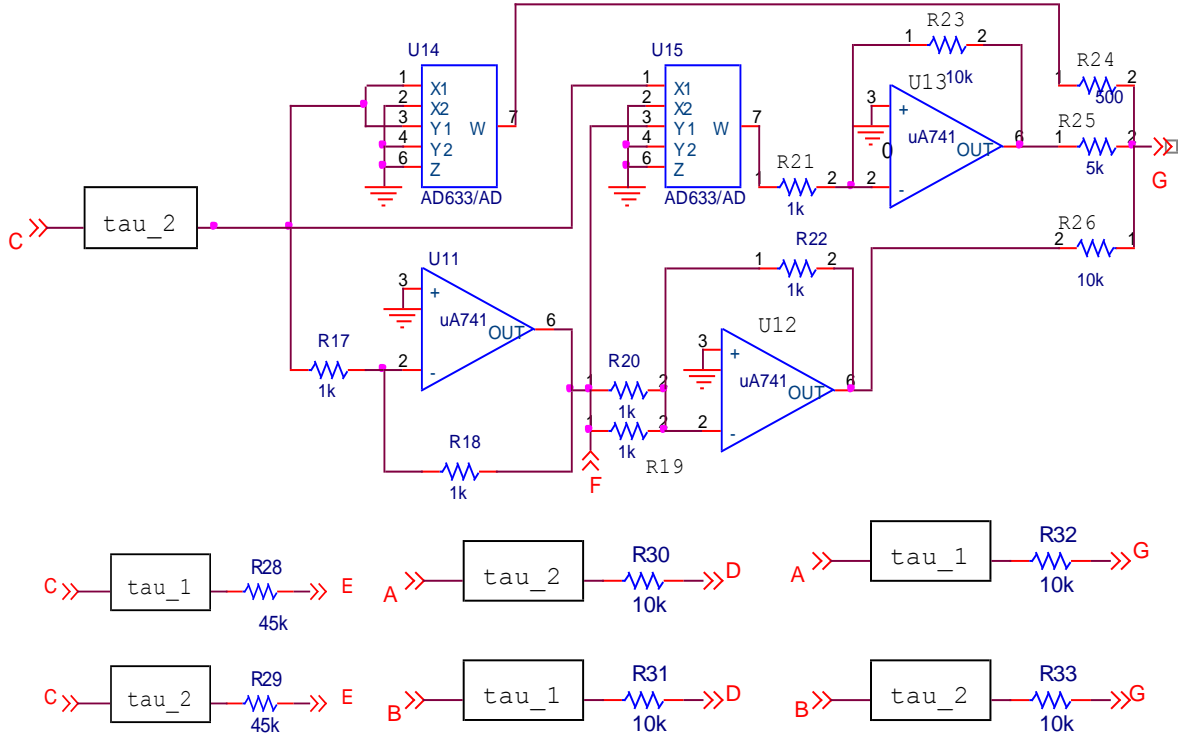

FIG. 4: Delay coupling circuit of the Sprott system. Each of the circuit for tau_i (i=1, 2, 3) is a LC ladder network where inductor L and capacitor C are appropriately chosen to design a desired delay τ.

## V. Conclusion

We explored a design of delay coupling for targeting multiple delays in two chaotic systems. We introduce different delays in different pairs of state variables of a drive-response system. The stability condition for this multiple lag configuration is derived with the help of the Hurwitz matrix stability criterion. Basically we designed the delay coupling for a driver oscillator where different delays are introduced in the coupling function and retrieved the delays at a response system. The main benefit is that one can communicate different information bits in the form of different lag configurations or delays between two distant oscillators. The parameter mismatch and the amount of delay does not affect the stability condition. We supported the theory of multiple delay with numerical simulations of the mismatched Rössler system and the slow-fast system Hindmarsh-Rose neuron model. We physically implemented the multiple delay configurations in electronic circuit of two coupled Sprott systems. To our best search of literature, we did not find any such example of multiple lag configurations.

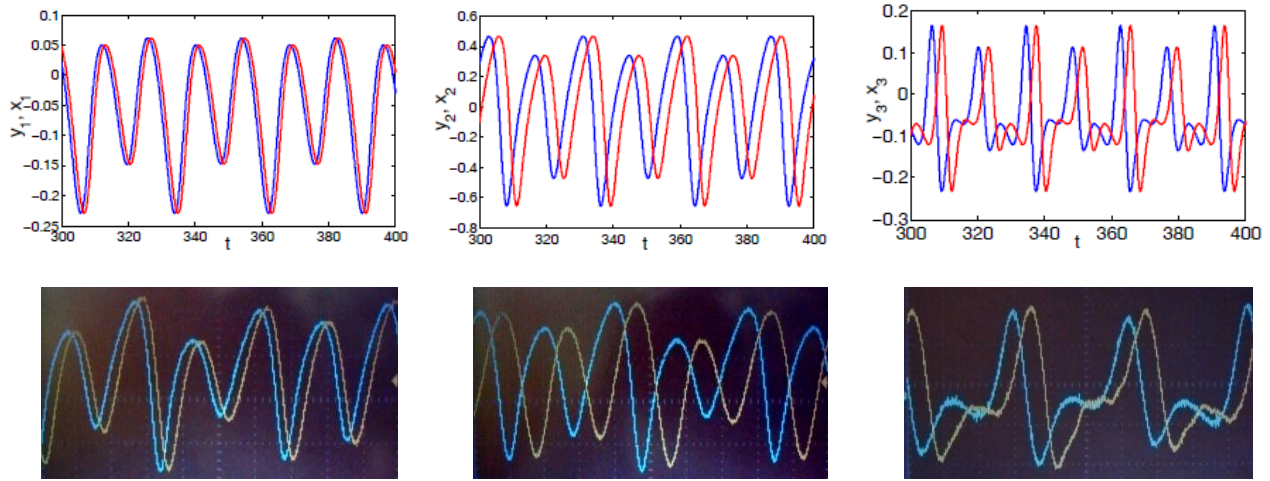

FIG. 5: Numerical and experimental (Oscilloscope picture) time series of the MLS in the upper and lower row. Here $\tau_1$=150 µs and $\tau_2=\tau_3$= 400 µs for experimental observations and $\tau_1$=1 and $\tau_2=\tau_3$=3 for numerical simulations.

**Acknowledgements**

S.K.B acknowledges support by the BRNS/DAE, India (Project No. 2009/34/26/BRNS). S.K.D. is supported by the CSIR Emeritus scientist scheme.